\documentclass[aps,twocolumn,groupedaddress,showpacs]{revtex4}
\usepackage{graphics}
\begin{document}
\newcommand{\beq}{\begin{equation}}
\newcommand{\eeq}{\end{equation}}

\title{Modified Two-Slit Experiments and Complementarity}

\author{Tabish Qureshi}
\affiliation{Centre for Theoretical Physics\\ Jamia Millia Islamia,
New Delhi-110025, India.\\
Email: tabish@ctp-jamia.res.in}


\begin{abstract}
Some modified two-slit interference experiments claim to demonstrate a
violation of Bohr's complementarity principle.  A typical such
experiment is theoretically analyzed using wave-packet dynamics. The
flaw in the analysis of such experiments is pointed out and it is
demonstrated that they do not violate complementarity. In addition, it is
quite generally proved that if the state of a particle
is such that the modulus square of the wave-function yields an interference
pattern, then it necessarily loses which-path information.
\end{abstract}


\maketitle

\section{Introduction}

It is generally accepted that the wave and particle
aspects of quantum systems are mutually exclusive. Niels Bohr elevated this
concept, which is probably born out of the uncertainty principle, to the
status of a separate principle, the principle of complementarity
\cite{bohr}.

Since then, the complementarity principle has been demonstrated in various
experiments, the most common of which is the two-slit interference
experiment. It has been shown that if the which-way information in a
double-slit experiment is stored somewhere, the interference pattern is
destroyed, and if one chooses to ``erase" the
which-way information {\em after} detecting the particle, the interference
pattern comes back. This phenomenon goes by the name of quantum erasure
\cite{eraser, eraser1}.

Recently, some interesting experiments were proposed and carried out which
claim to violate the complementarity principle \cite{chown,afshar,afsharfp}.
A schematic diagram of a typical such experiment is shown in Fig. 1.
Basically, it
consists of a standard two-slit experiment, with a converging lens $L$ behind
the conventional screen for obtaining the interference pattern. Although
the experiments use pinholes instead of slits, we will continue to refer to
them as slits.
If the screen is removed, the light passes through the lens and produces
two images of the slits, which are captured on two detectors $D_A$ and
$D_B$ respectively. Opening only slit $A$ results in only detector $D_A$
clicking, and opening only slit $B$ leads to only $D_B$ clicking. The
authors argue that the detectors $D_A$ and $D_B$ yield information about which
slit, $A$ or $B$, the particle initially passed through. If one places a
screen before the lens, the interference pattern is visible.

Conventionally, if one tries to observe the interference pattern, one
cannot get the which-way information. The authors have a clever scheme for
establishing the existence of the interference pattern without actually
observing it. First the exact location of the dark fringes are noted by
observing the interference pattern. Then, thin wires are placed in the
exact locations of the dark fringes. The argument is that if the
interference pattern exists, sliding in wires through the dark fringes will
not affect the intensity of light on the two detectors. If the interference
pattern is not there, some photons are bound to hit the wires, and get
scattered, thus reducing the photon count at the two detectors. This way,
the existence of the interference pattern can be established without
actually disturbing the photons in any way. This is similar in spirit to
the so called ``interaction-free measurements" where the non-observation of
a particle along one path establishes that it followed the other possible
path, without actually measuring it \cite{kwait}. The authors carried out the
experiment and found that sliding in wires in the expected locations of the
dark fringes, doesn't lead to any significant reduction of intensity at the
detectors. Hence they claim that they demonstrated a violation of
complementarity.

\begin{figure}
\centerline{\resizebox{8.0cm}{!}{\includegraphics{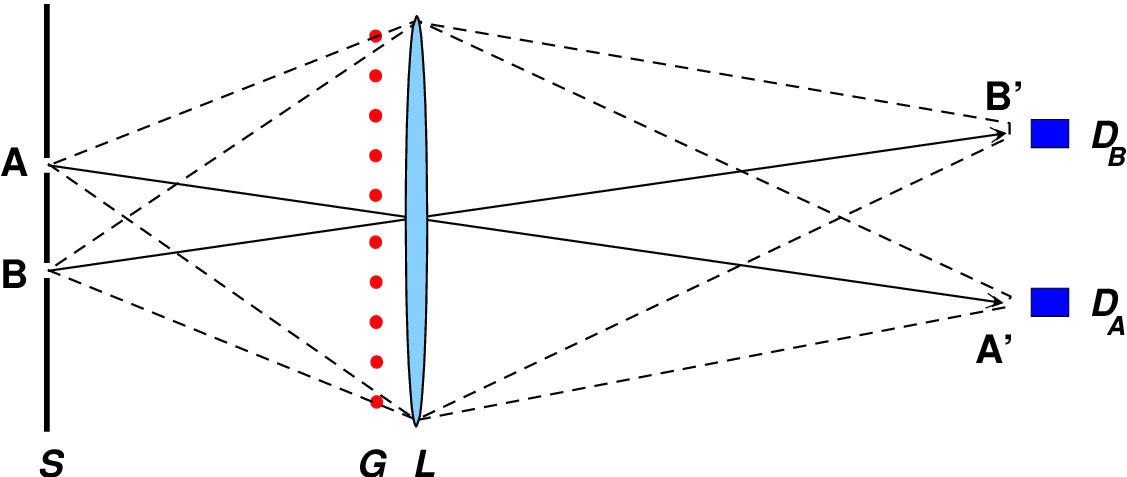}}}
\caption{ A schematic representation of the modified two-slit experiment. Light emerges
from two pin-holes (not slits) A and B and interferes. Thin wires are
placed carefully in the exact locations of the dark fringes of the interference
pattern. The lens L collects the light and obtains the images of the two
slits A' and B', respectively. The detectors $D_A$ and $D_B$ collect the
photons for these two images.}
\end{figure}

The complementarity principle is at the heart of quantum mechanics, and its
violation will be deeply disturbing to its established understanding.
As expected, there has been skepticism towards the modified two-slit
experiment, and a
heated debate followed \cite{unruh, kastner, kastner1, tq, srinivasan, jacques, drezet, steuernagel, ohara}.
Interestingly, different criticisms do not agree with each other on why
complementarity is not violated in this experiment. 
None of the criticisms has been
satisfactorily able to point out any flaw in the interpretation of the
experiments. Most agree that if the introduction of wires has no effect on
the intensity, it shows that interference exists.
Almost everybody seems to agree that detecting a photon
at (say) $D_A$ means that it came from slit A. However,
just because blocking slit B leads to only detector $D_A$
clicking and blocking slit A leads to only detector $D_B$
clicking, quantum mechanics doesn't say that when both
slits are open, detector $D_A$ clicking implies that the
photon came from slit A. This has also been pointed out
by Kastner\cite{kastner}.

Some authors\cite{jacques,steuernagel} have tried to argue along the
following line. Blocking dark fringes also blocks out
parts of the bright fringes. They believe, when both slits
are open, the photons contain full which-slit information,
which is partially destroyed by partially blocking the
bright fringes. They argue that since without the wires
completely blocking the dark fringe, one cannot infer the
existence of the interference pattern, when one tries to
increase the information about the existence of the interference,
the which-slit information is proportionately
decreased. However, in these works, the {\em calculation}
shows the existence of full interference, without any
blocking wires. Without the blocking wires, the photons
are claimed to have full which-slit information (distin-
guishability = 1, in their language). So, if one were to
take their calculation of distinguishablity as correct, then
the existence of full interference {\em in the calculation} (in the
sense of $|\psi(x)|^2$ yielding an interference pattern) seems
to imply that quantum formalism allows existence of
interference pattern for photons for which full which-slit
information exists. This clearly goes against the spirit of
complementarity. In our view, the calculation of distinguishability
in these works is fundamentally flawed.

Many authors have have fallen
back to a more formal interpretation of Bohr's principle, namely that the
wave and particle nature cannot be seen in the same experiment, for the
same photons. They argue that authors have to do not one, but two experiments
to prove their point - one without the wires, one with the wires. Some argue
that the existence of fringes has already been assumed, and that the argument
of the experiment is circular. However, a reader who respects empirical
facts, doesn't see it
as two experiments if putting in the wires is not changing the results. One
would say, the interference is out there in the middle, and one can check 
out that the photons are not passing through certain regions, the dark
fringes.

Let us understand what is happening in the experiment slightly better by
simplifying the experiment. One might argue that a two-slit experiment is
the simplest experiment one can imagine. But this experiment still has a large
number of degrees of freedom, and the Hilbert space is big. We will show in
the following that the simplest interference (thought) experiment can be just
carried out using a spin-1/2 particle.

\section{A simplified ``two-slit" experiment}

Let there be a spin-1/2 particle traveling along x-axis. For our purpose,
its physical motion is unimportant - we will only be interested in the
dynamics of its spin. In the two-slit experiment, the particle emerges
from the slits in a superposition of two physically separated, localized
wave-packets, which are orthogonal to each other. Their physical separation
guarantees their orthogonality. Any subsequent unitary evolution will
retain their orthogonality. We assume the initial state of the spin to be
\begin{equation}
|\psi\rangle_0 = {1\over\sqrt{2}}(|\uparrow\rangle + |\downarrow\rangle) .
\label{psi0}
\end{equation}
Here, the states $|\uparrow\rangle$ and $|\downarrow\rangle$ are the
eigenstates of $\hat{S}_z$, and play the role
of the two wave-packets that emerge from the double-slit. The time
evolution, in the conventional two-slit experiment, spreads the
wave-packets so that they overlap. In our thought experiment, we employ a
homogeneous magnetic field $B$, acting along the y-axis, to evolve
the two states $|\uparrow\rangle$ and $|\downarrow\rangle$. Thus the
Hamiltonian of the system is $\hat{H} = B\hat{S}_y$.

The time evolution operator can be written as 
\begin{eqnarray}
\hat{U}(t) &=& \exp({i\over\hbar}B\hat{S}_yt)\nonumber\\
	   &=& \cos(Bt/2) + i\hat{\sigma}_y\sin(Bt/2),
\end{eqnarray}
where $\hat{\sigma}_y$ is the usual Pauli matrix. The time evolution
operator, for  a time $\tau= \pi/2B$ has the form
\begin{equation}
\hat{U}(\tau) = {1\over\sqrt{2}} (1 + i\hat{\sigma}_y) .
\end{equation}
It is straight forward to see that the time evolution transforms the
states$|\uparrow\rangle$ and $|\downarrow\rangle$ as
\begin{eqnarray}
\hat{U}(\tau)|\uparrow\rangle &=&  {1\over\sqrt{2}}(|\uparrow\rangle -
|\downarrow\rangle)\nonumber\\
\hat{U}(\tau)|\downarrow\rangle &=&  {1\over\sqrt{2}}(|\uparrow\rangle +
|\downarrow\rangle)
\end{eqnarray}
A further evolution through a time $\tau$ will transform the states as
\begin{eqnarray}
\hat{U}(2\tau)|\uparrow\rangle &=&  - |\downarrow\rangle \nonumber\\
\hat{U}(2\tau)|\downarrow\rangle &=&  |\uparrow\rangle \label{spinend}
\end{eqnarray}
After an evolution through a total time $2\tau$, if one puts a spin
detector, one will either get a ``spin-down" or a ``spin-up". In the
beginning, if we started out with a state $|\uparrow\rangle$, the detector
at the end will register a $|\downarrow\rangle$ state. On the other hand, of
we started with a $|\downarrow\rangle$ state, the detector
at the end will register a $|\uparrow\rangle$ state. Thus, the detector at
the end obtains a which-initial-state information about the spin, exactly as
the detectors in the modified two-slit experiment obtain a which-slit information.

So, now we carry out our thought experiment, with the initial state
$|\psi\rangle_0$, as given by (\ref{psi0}). After a time $\tau$ we get into
a state which is equivalent to the interference region in the modified two-slit
experiment:
\begin{equation}
\hat{U}(\tau)|\psi\rangle_0 =  {1\over 2}(|\uparrow\rangle -
|\downarrow\rangle) + {1\over 2}(|\uparrow\rangle + |\downarrow\rangle) .
\label{interf}
\end{equation}
Equation (\ref{interf}) represents an interference pattern, because the
``down-spins" cancel out to give destructive interference, while the
``up-spins" add up to give constructive interference. So, in our two-state
interference experiment, there is one dark fringe and one bright fringe.
After parts of the states have been destroyed by the ``dark-fringe", what
is left is just
\begin{equation}
\hat{U}(\tau)|\psi\rangle_0 =  {1\over 2}|\uparrow\rangle 
+ {1\over 2}|\uparrow\rangle .
\label{interf1}
\end{equation}

Now, these two contribution from the two initially orthogonal states are
identical. States which are not orthogonal, are naturally not
distinguishable. Thus the which-way or which-initial-state information is
lost at this stage. It might serve some useful purpose by keeping the two
contributions separate and evolving them for a further time $\tau$ to see
what they yield at the spin detector.
\begin{equation}
\hat{U}(\tau)({1\over 2}|\uparrow\rangle 
+ {1\over 2}|\uparrow\rangle) =  {1\over 2\sqrt{2}}(|\uparrow\rangle -
|\downarrow\rangle) + {1\over 2\sqrt{2}}(|\uparrow\rangle - |\downarrow\rangle) .
\end{equation}
So, we see that each part of the initial state
leads to a superposition of $|\downarrow\rangle$ and $|\uparrow\rangle$
states. So, although the spin-detector at the end still gives either a
``spin-up" or a ``spin-down", it gives no information about whether the
initial state was a $|\downarrow\rangle$ or $|\uparrow\rangle$. Yet, if one
started out with either a $|\downarrow\rangle$ or $|\uparrow\rangle$ initial
state, which is equivalent of closing one slit, (\ref{spinend}) tell us that
each will go to only its corresponding detector, thus giving the
which-initial-state information.

Taken in isolation, this over-simplified thought experiment may not
prove anything substantial, but it does provide a clue to where the problem
lies in the modified two-slit experiments. It appears that the very existence of
interference destroys which-way information. 

\section{Interference and which-way information}

\subsection{Which-way information destroys interference}

There have been some recent developments in understanding the origins of
complementarity. It has been argued, and demonstrated, that one need not
actually carry out a which-way measurement in order that the
interference disappears. Mere existence of which-way information in the state is
sufficient to destroy any potential interference \cite{eraser1, durr}.
In other words, one can say that mere
existence of the possibility of getting which-way information is enough
to destroy the interference pattern. This can be simply demonstrated as
follows. Suppose that the state of a particle having passed through a
double-slit, just before hitting the screen, is given by
\begin{equation}
             \psi(x) = \psi_A(x) + \psi_B(x)
\end{equation}
where $\psi_A(x)$ and $\psi_B(x)$ represent the amplitude of the particle
passing through slit 1 and 2, respectively. Probability of finding the
particle at a point x on the screen, is given by
\begin{equation}
 |\psi(x)|^2 = |\psi_A(x)|^2 + |\psi_B(x)|^2 +
               \psi_A^*(x)\psi_B(x) + \psi_B^*(x)\psi_A(x)
\end{equation}
The last two terms represent interference. Now, let us suppose that
in a slightly modified version of this experiment, the state is given by
\begin{equation}
             \psi(x) = \psi_A(x)|1\rangle + \psi_B(x)|2\rangle
\end{equation}
where $|1\rangle$ and $|2\rangle$ are certain othonormal
eigenstates of a suitable observable, which get entangled with the particle,
and hence carry which-way
information. A measurement of that observable yielding $|1\rangle$ will
lead to a definitive conclusion that the particle passed through slit 1,
and likewise for $|2\rangle$. In this case
the probability of finding the particle at a point x, is given by
\begin{equation}
     |\psi(x)|^2 = |\psi_A(x)|^2 + |\psi_B(x)|^2
\end{equation}
The two terms which would have given interference, are killed by the
orthogonality of $|1\rangle$ and $|2\rangle$. One would notice that a
which-way measurement
is not even needed here - mere existence of which-way information,
or mere possibility of a which-way measurement, is enough to destroy
interference. 

\subsection{Interference destroys which-way information}

In the following we will prove that the converse of {\em which-way
information necessarily destroys interference} is also true. This would
mean that a quantum state which has a form required to yield interference,
cannot contain any which-way information.

In any variant of the conventional two-slit experiment, the
initial state has to be in a superposition of two orthogonal states. This
follows simply from the fact that the two slits are distinguishable. In
the case of a Mach-Zehnder interferometer, a half-silvered mirror splits
the beam into superposition of two spatially separated beams. Thus
the initial (unnormalized) state $|\psi\rangle_0$ can be written as
\begin{equation}
|\psi\rangle_0 = |\psi_A\rangle_0 + |\psi_B\rangle_0 ,
\end{equation}
These states then evolve and reach the region where they spatially overlap.
$|\psi_A\rangle_0$ and $|\psi_B\rangle_0$ evolve to $|\psi_A\rangle$ and
$|\psi_B\rangle$ respectively, so that in the region of overlap, the
state looks like
\begin{equation}
|\psi\rangle = |\psi_A\rangle + |\psi_B\rangle .
\label{initialstate}
\end{equation}
 The time evolution being unitary, these two
parts retain their orthogonality, so that $\langle\psi_A|\psi_B\rangle = 0$.
Let us assume that in the region where they interfere, they have the form:
\begin{eqnarray}
|\psi_A\rangle &=& |\alpha\rangle + |\gamma\rangle \nonumber\\
|\psi_B\rangle &=& |\beta\rangle - |\gamma\rangle ,
\label{state}
\end{eqnarray}
where $|\gamma\rangle$ is chosen to be normalized, keeping
$|\psi_A\rangle$ and $|\psi_B\rangle$ unnormalized, which can always be done.
Here, $|\alpha\rangle$, $|\beta\rangle$ and $|\gamma\rangle$ need not be
simple states - each may involve a multitude of degrees of freedom.
However, $|\psi_A\rangle$ and $|\psi_B\rangle$ have to have this form,
so that there are parts from the two which
cancel out exactly, to give the so-called dark-fringes. 
In the case of the conventional two-slit experiment, $|\gamma\rangle$
would constitute the pattern involving all the dark fringes (a specific
example of this appears in the next section). 
For the case of a Mach-Zehnder interferometer,
$|\gamma\rangle$ represents the part of the state from one beam, reaching one
of the two detectors. Interference happens when $|\gamma\rangle$ parts reaching one
detector, from both the beams, cancel out. The result is, that particular
detector does not detect any particles (dark fringe).
Thus, if a state described by (\ref{initialstate},\ref{state}) has a non-zero
$|\gamma\rangle$, one can be sure that a measurement will lead to an
interference pattern. One can thus associate a non-zero $|\gamma\rangle$
with the existence of interference, even without doing an actual measurement.

If complementary could indeed be violated, the parts of the two initial
states which are left (after the destructive interference),
namely $|\alpha\rangle$ and $|\beta\rangle$, should be
orthogonal, so that they can contain a which-way information. States which
are not orthogonal, cannot lead to distinguishable outcomes in a measurement.
In order that $\gamma$ describes the pattern of dark fringes, it should
be orthogonal to $|\alpha\rangle$ and $|\beta\rangle$, so that it is 
distinguishable from the other parts. However, the following analysis
goes through even without that restriction.

The norm of $|\psi\rangle$ is given by 
\begin{eqnarray}
\langle\psi|\psi\rangle &=&
\langle\psi_A|\psi_A\rangle + \langle\psi_B|\psi_B\rangle \nonumber\\
&=& \langle\alpha|\alpha\rangle + \langle\alpha|\gamma\rangle +
\langle\gamma|\alpha\rangle +\langle\gamma|\gamma\rangle \nonumber\\
&& + \langle\beta|\beta\rangle - \langle\beta|\gamma\rangle -
\langle\gamma|\beta\rangle +\langle\gamma|\gamma\rangle \nonumber\\
&=& \langle\alpha|\alpha\rangle + \langle\alpha|\gamma\rangle +
\langle\gamma|\alpha\rangle \nonumber\\
&& + \langle\beta|\beta\rangle - \langle\beta|\gamma\rangle -
\langle\gamma|\beta\rangle + 2 \label{norm1}
\end{eqnarray}
Using orthogonality of $|\psi_A\rangle$ and $|\psi_B\rangle$, we get
\begin{equation}
\langle\alpha|\gamma\rangle - \langle\gamma|\beta\rangle =
1 - \langle\alpha|\beta\rangle .
\end{equation}
Substituting this in (\ref{norm1}), yields
\begin{equation}
\langle\psi|\psi\rangle =
\langle\alpha|\alpha\rangle + \langle\beta|\beta\rangle -
\langle\alpha|\beta\rangle - \langle\beta|\alpha\rangle + 4\label{norm1a}
\end{equation}
In the region of overlap the state has the actual form
 $|\psi\rangle = |\alpha\rangle + |\beta\rangle$. Using this, the norm of
$|\psi\rangle$ can be written as 
\begin{equation}
\langle\psi|\psi\rangle = \langle\alpha|\alpha\rangle +
\langle\beta|\beta\rangle + \langle\alpha|\beta\rangle +
\langle\beta|\alpha\rangle \label{norm2}
\end{equation}

Combining (\ref{norm1a}) and (\ref{norm2}), we get
\begin{equation}
\langle\alpha|\beta\rangle + \langle\beta|\alpha\rangle = 2
\end{equation}
From the above equation it is obvious that $|\alpha\rangle$ and
$|\beta\rangle$ can never be orthogonal. Hence $|\psi\rangle$ in
overlap region contains no which-way information.

This general analysis shows that if the state has a form
which could yield interference, it cannot contain any which-way information.
In other words, existence of interference necessarily destroys which-way
information.

So, complementarity is robust, and cannot be violated in any such interference
experiment where one tries to look for which-way information after interference.

\section{Two-slit experiment with wave-packets}

Coming back to the modified two-slit experiment, let us see what implication the
preceding analysis has on it. Clearly, in the modified two-slit experiment, after the
particle passes through the interference region, the which way information is
lost. The detector $D_A$ clicking doesn't mean that the particle came from
slit $A$. It might appear hard to visualize that a wave-packet which
travels in a straight line from slit $A$ can contribute to a click on detector
$D_B$, which is not in its direct path. However, the argument of momentum
conservation will hold only if we {\em knew} that
the particle started out from slit $A$ - in that situation it would
never reach detector $D_B$. In order to demonstrate the validity of these
arguments for the modified two-slit experiment, let us look at the two-slit experiment
in more detail.

We carry out the analysis for massive particles, to show that the argument
doesn't just hold for photons. Consider the particle to be
moving along the x-axis, and the slit plane to be parallel to the y-axis.
We will only be interested in the dynamics of the state in the y-direction,
whereas the x-axis motion just serves the purpose of transporting the
particle from the slits to the detectors.
Let us assume that when the particle emerges from the double-slit, its state
is given by  a superposition of two distinct, spatially localized
wave-packets. The state at this time, which we choose to call $t=0$, can be
written as
\begin{equation}
\psi(y,0) = {a\over(\pi/2)^{1\over 4}\sqrt{\epsilon}} e^{-(y-y_0)^2/\epsilon^2}
+ {b\over(\pi/2)^{1\over 4}\sqrt{\epsilon}} e^{-(y+y_0)^2/\epsilon^2}
\label{wpsi0}
\end{equation}
where $\epsilon$ is the width of the wave-packets, $2y_0$ is the slit
separation, and $a$ and $b$ are the amplitudes of the two wave-packets.

The wave-packets evolve in time, during which they spread, and reach the
region where they overlap, and thus, interfere. This state can just be
obtained by evolving (\ref{wpsi0}), using the Hamiltonian
$\hat{H}=\hat{p}_y^2/2m$. Hence, the state at time $t$ can be written as
\begin{eqnarray}
\psi(y,t) &=& 
a C(t) e^{-{(y-y_0)^2\over\epsilon^2+2i\hbar t/m}}
+ b C(t) e^{-{(y+y_0)^2\over\epsilon^2+2i\hbar t/m}}
\end{eqnarray}
where 
\begin{eqnarray}
C(t) &=& {1\over (\pi/2)^{1/4}\sqrt{\epsilon+2i\hbar
t/m\epsilon}}.
\end{eqnarray}
Now, this state can also be rewritten as
\begin{eqnarray}
\psi(y,t) &=& 
a C(t)e^{-{y^2+y_0^2\over\Omega(t)}}
\left( \cosh({2yy_0\over\Omega(t)}) +
\sinh({2yy_0\over\Omega(t)})\right)\nonumber\\
&&+ b C(t)e^{-{y^2+y_0^2\over\Omega(t)}}
\left( \cosh({2yy_0\over\Omega(t)}) -
\sinh({2yy_0\over\Omega(t)})\right) ,\nonumber\\
\end{eqnarray}
where $\Omega(t) = \epsilon^2+2i\hbar t/m$.
The term involving $\sinh({2yy_0\over\Omega(t)})$ in this expression 
is an example of $|\gamma\rangle$ introduced in the preceding section.

In the usual case of a two-slit experiment, the amplitudes coming from the
two slits are approximately equal, i.e., $a=b=1/\sqrt{2}$. In this case, the
sinh terms cancel out to give the dark fringes, and what is left is
\begin{eqnarray}
\psi(y,t) &=& 
{1\over 2}a C(t)\left( e^{-{(y-y_0)^2\over\Omega(t)}} +
e^{-{(y+y_0)^2\over\Omega(t)}}\right) \nonumber\\
&&+ {1\over 2}b C(t)\left( e^{-{(y-y_0)^2\over\Omega(t)}} +
e^{-{(y+y_0)^2\over\Omega(t)}}\right) . \label{psit}
\end{eqnarray}
In this state, the parts of the state coming from the two slits
are equal. So, there is no which-way information in the state any more.
If a lens is used after this stage, which takes the part
$e^{-{(y-y_0)^2/\Omega(t)}}$ to one detector and the part
$e^{-{(y+y_0)^2/\Omega(t)}}$ to the other detector, one can easily see from
(\ref{psit}) that a part coming from one slit, becomes a superposition of
two parts going to the two detectors. So, each detectors receives equal
contribution from the two slits, and registering a particle gives no
information on which slit the particle came from. To a mind used to 
classical way of thinking, it might appear that the particle received at
a particular detector came from a particular slit, but the preceding analysis
shows that this presumption is erroneous.

Note however, that if one of the slits is closed, say, if $a=1$ and $b=0$,
the state at time $t$ will be $a C(t)e^{-{(y-y_0)^2\over\Omega(t)}}$, which
goes to only one detector when a lens is used.

A critic might argue that instead of canceling out some parts of the two
wave-packets, one might just evolve them separately, and because they were
initially orthogonal, they will give distinct result at the end. The answer
to that is, if one really looks for interference by putting thin wires, one
is blocking those very parts of the wave-packets which are canceling out.
So, those parts will not reach the detectors, even if one insists on not
canceling them out. On the other hand, the argument of bringing in wires is
not really needed. The state (\ref{psit}) as such, has no which path
information.

\section{Conclusions}

Although the modified two-slit experiments do have genuine
interference, as shown by introducing thin wires, the detectors detecting the
photons behind the converging lens, do not yield any which-way information.
It was not easy to see this without mathematically rigor ous arguments.
The wave-packet analysis makes this fact trivially obvious. 
Many earlier analysis of complementarity have concentrated on showing how
existence of which-way information destroys interference. We have taken the
reverse approach, as demanded by the the modified two-slit experiment.
We have shown that if a state is such that the modulus square of itsa
wave-function yields and interference pattern, it cannot contain any
which-way information.
The complementarity principle is thus robust and cannot
be violated in any experiment which is a variant of the two-slit experiment.
We see yet again that in dealing with quantum systems, trusting classical
intuition can easily lead one to wrong conclusions.

\section{Acknowledgments}
The author thanks R. Srikanth for very useful discussions.
The author thanks the Centre for Philosophy and Foundations of Science,
New Delhi, for organizing annual meetings on foundations of quantum
mechanics, which provide a platform for exchange of ideas.

\end{document}